\documentstyle[12pt]{article}
\newcommand{\be}{\begin{equation}}
\newcommand{\ee}{\end{equation}}
\newcommand{\bea}{\begin{eqnarray}}
\newcommand{\eea}{\end{eqnarray}}
\newcommand{\ba}{\begin{array}}
\newcommand{\ea}{\end{array}}

\newcommand{\norsl}{\normalsize\sl}
\newcommand{\norsc}{\normalsize\sc}
\begin{document}


\begin{titlepage}

\title{ 
\vskip -3cm
{\normalsize
\begin{flushright}
KUCP-102   \\
October, 1996
\end{flushright}
}
\vskip 3cm
 Spectral Flow and Feigin-Fuks Parameter Space 
         of N=4 Superconformal Algebras}

\author{
\vspace*{1cm}\\
\norsc  Satoshi MATSUDA\thanks{Work supported in part by
          the {\it Monbusho} Grant-in-Aid for Scientific Research
          on Priority Areas 231 ``Infinite Analysis", 
          No. 08211227.} 
         \thanks{e-mail address:
           matsuda@phys.h.kyoto-u.ac.jp}\\
\vspace{-3mm}\\
\norsl  Department of Fundamental Sciences\\
\norsl  FIHS, KYOTO UNIVERSITY\\
\norsl  Kyoto 606-01, JAPAN}

\date{}
\maketitle

\vspace{2cm}

\begin{abstract}
{\normalsize
The parameter space of the Feigin-Fuks representations of 
the N=4  SU(2)$_k$ superconformal algebras is studied 
from the viewpoint of the specral flow. 
The $\eta$ phase of the spectral flow is nicely incorporated 
through twisted fermions 
and the spectral flow resulting from 
the inner automorphism of the N=4 superconformal algebras
is explicitly shown to be operating as identiy relations 
among the generators. 
Conditions for the unitary representations are also investigated 
in our Feigin-Fuks parameter space.  
}
\end{abstract}

\begin{picture}(5,2)(-300,-615)
\put(2.3,-110){}
\put(2.3,-125){}
\put(2.3,-140){}
\put(2.3,-155){}
\end{picture}  

\thispagestyle{empty}
\end{titlepage}

\setcounter{page}{1}
\baselineskip 24pt
It is well recognized nowadays that the so-called Feigin-Fuks (FF)
representations (or the Coulomb-gas representations)~\cite{FF,DotseF} 
are very important and almost inevitably required tools for 
investigating the representation theories of the conformal 
and superconformal algebras.
By now we have established the FF representations of the 
superconformal algebras with higher number of supercharges
~\cite{KatoM1/2,KatoM3,KatoMR,Miki}, 
up to N=4~\cite{Mat1,Mat2,ItoMP}. 

On the other hand, the spectral flows resulting from 
the inner automprphisms of 
the conformal and superconformal algebras 
with N=2,3 and 4 were first recognized 
by Schwimmer and Seiberg~\cite{SchwS}, 
and their remarkable implications on the representation theories of 
the algebras have been discussed by many people
~\cite{EguchiT3,DefeverST,PetersenT}.

In the present paper we shall study 
the parameter space of the FF representations 
of the N=4 SU(2)$_k$ superconformal algebras 
with particular focus on the properties of their spectral flow 
which is nicely embedded in our FF parameterization~\cite{Mat1}. 
We shall explicitly show how remarkably the spectral flow emerges 
by use of the identities holding among the FF parameters.
Our study not only shows how the spectral flow for the 
unitary representations~\cite{PetersenT,EguchiT12,Yu} 
of the N=4 SU(2)$_k$ superconformal algebras is operating, 
but also establishes it to hold explicitly 
in the nonunitary representations 
by use of the continuous parameters of our FF representations~\cite{Mat1}. 

The N=4 SU(2)$_k$ superconformal algebras are defined by the 
form of the operator product expansions (OPE) among operators 
given by the energy-momentum tensor $L(z)$, 
the SU(2)$_k$ local nonabelian generators $T^i(z)$, and 
the iso-doublet and -antidoublet supercharges $G^a(z)$ and $\bar{G}_a(z)$:
\begin{eqnarray}
  L(z)L(w) &\sim& {3k\over (z-w)^4}+{2L(w)\over(z-w)^2}+
  {\partial_wL(w)\over z-w},                                \nonumber \\
  T^i(z)T^j(w) &\sim& {{1\over2}k\eta^{ij}\over(z-w)^2}+
      {{\rm i}\epsilon^{ijk}\eta_{kl}T^l(w)\over z-w},\ 
   \ L(z)T^i(w)\sim {T^i(w)\over (z-w)^2}+
        {\partial_wT^i(w)\over z-w},                         \nonumber \\
  T^i(z)G^a(w) &\sim& -{{1\over 2}{(\sigma^i)^a}_bG^b(w)\over z-w}\ ,\quad 
    \qquad  T^i(z)\bar G_a(w)\sim 
      {{1\over 2}\bar G_b(w){(\sigma^i)^b}_a\over z-w}\ ,     \nonumber \\
  L(z)G^a(w) &\sim& {{3\over2}G^a(w)\over(z-w)^2}+
     {\partial_w G^a(w)\over z-w}\ ,\ 
    \ L(z)\bar G_a(w)\sim{{3\over2}\bar G_a(w)\over(z-w)^2}+
         {\partial_w \bar G_a(w)\over z-w}\ ,                 \nonumber \\
  G^a(z)G^b(w) &\sim& 0\ ,\quad \bar G_a(z)\bar G_b(w)\sim 0, \nonumber \\
  G^a(z)\bar G_b(w) &\sim& {4k{\delta^a}_b\over(z-w)^3}-
     {4{(\sigma^i)^a}_b{\eta_{ij}}T^j(w)\over(z-w)^2}-
     {2{(\sigma^i)^a}_b{\eta_{ij}}\partial_w T^j(w)\over z-w}+
     {2{\delta^a}_bL(w)\over z-w} 
  \label{OPE}
\end{eqnarray}
where $i=(\pm,0)$ denote SU(2) triplets in the diagonal basis, while
the superscripts (subscripts) $a=(1,2)$ label SU(2) doublet (antidoublet) 
representations.
The symmetric tensors $\eta^{ij}=\eta^{ji}=\eta_{ij}$ 
are defined in the diagonal basis as 
$\eta^{+-}=\eta^{00}=1$, while the antisymmetric tensors 
$\epsilon^{ijk}=-\epsilon^{jik}=-\epsilon^{ikj}$ are 
similarly defined as $\epsilon^{+-0}=-{\rm i}$, etc., 
and otherwise zero.
The Pauli matrices are given by 
$\sigma^{\pm}=(\sigma^1\pm{\rm i}\sigma^2)/\sqrt 2, \sigma^0=\sigma^3$. 
The corresponding notations 
in terms of the isospin raising and lowering by a half unit
for the fermionic operators 
are given by 
$G^1\equiv G^-,\, G^2\equiv G^+,\, 
\bar G_1\equiv \bar G_+,\, \bar G_2\equiv \bar G_-$.
The symmetric delta function ${\delta^a}_b={\delta^b}_a$ 
has the standard meaning with 
${\delta^-}_+={\delta^1}_1=1,\, {\delta^+}_+={\delta^2}_1=0, etc.$.

Here is a remark in order. 
The $N=4$ algebras Eq.(\ref{OPE}) have a global SU(2) symmetry. 
As a result, one can consider the infinite number of 
independent $N=4$ 
twisted algebras labeled by $\rho$ corresponding to the 
conjugate classes of the global automorphism\cite{SchwS}. 
The FF representations of the $\rho$-extended $N=4$ algebras 
are discussed in a separate paper\cite{MatIshimoto}. 
The present paper will be restricted to the case of $\rho=0$.

As noted by Schwimmer and Seiberg~\cite{SchwS}, 
the SU(2) gauge symmetry gives 
the {\it inner} automorphism 
\bea
L(z) &\rightarrow& L(z)+{\rm i}{d\alpha(z)\over dz}T^0(z)
-{k\over 4}\left({d\alpha(z)\over dz}\right)^2, \nonumber \\
T^0(z) &\rightarrow& T^0(z)+{\rm i}{k\over 2}{d\alpha(z)\over dz},\quad
T^{\pm}(z) \rightarrow e^{\pm{\rm i}\alpha(z)}T^{\pm}(z), \nonumber \\
G^{\mp}(z) &\rightarrow& e^{\mp{\rm i}{\alpha(z)\over 2}}G^{\mp}(z),\quad
\bar G_{\pm}(z) \rightarrow 
e^{\pm{\rm i}{\alpha(z)\over 2}}\bar G_{\pm}(z) 
\label{Auto}
\eea
for the N=4 SU(2)$_k$ superconformal algebras, 
while the boundary conditions are imposed 
by use of one parameter $\eta$ as
\begin{eqnarray} 
T^{\pm}(z) &=& e^{\mp 2{\rm i}\pi\eta}T^{\pm}(e^{2{\rm i}\pi} z)\ ,
\nonumber\\  
G^{\mp}(z) &=& -e^{\pm {\rm i}\pi\eta}G^{\mp}(e^{2{\rm i}\pi}z)\ ,
\nonumber\\ 
\bar G_{\pm}(z) &=& -e^{\mp {\rm i}\pi\eta}\bar G_{\pm}(e^{2{\rm i}\pi}z)\ .
\label{BoundC}
\end{eqnarray}
This $\eta$ phase can be gauged away through the use of the 
local automorphism Eq.(\ref{Auto}) by the choice 
$\alpha(z)={\rm i}\eta\log z$~\cite{SchwS}.

For the convenience of our later use we shall give here the 
N=4 SU(2)$_k$ superconformal algebras in terms of Fourier components:
\begin{eqnarray}
  [L_m,L_n] &=& (m-n)L_{m+n}+{k\over 2}m(m^2-1)\delta_{m+n,0}, \nonumber\\~
  [T^i_m,T^j_n] &=& {\rm i}\epsilon^{ijk}\eta_{kl} T^l_{m+n}
           +{k\over 2}m\delta_{m+n,0}, \quad        
  [L_m,T^i_n]=-nT^i_{m+n},                          \nonumber\\~
  [T^i_m,G^a_r] &=& -{1\over 2}{(\sigma^i)^a}_b G^b_{m+r}, \quad 
  [T^i_m,{\bar G}_{a,r}]
        ={1\over 2}{\bar G}_{b,m+r}{(\sigma ^i)^b}_a, \nonumber \\~
  [L_m,G^a_r] &=&({1\over 2}m-r)G^a_{m+r},\quad
  [L_m,\bar G_{a,r}]=({1\over 2}m-r)\bar G_{a,m+r},  \nonumber \\~
  \{G^a_r,G^b_s\} &=&0,\quad   \{\bar G_{a,r},\bar G_{b,s}\}=0,  \nonumber\\~
  \{G^a_r,\bar G_{b,s}\} &=& 2{\delta^a}_b L_{r+s}
  -2(r-s){(\sigma^i)^a}_b \eta_{ij}T^j_{r+s} 
  +{k\over 2}(4r^2-1)\delta_{r+s,0}{\delta^a}_b
       \label{Fourier}
\end{eqnarray}
where the following Fourier modings are to be given to complete the 
definition of the algebras: 
$L_m,\, T^0_m\, (m\in Z),\, T^{\pm}_m\, (m\in Z\mp\eta), 
G^{\pm}_r,\, \bar G_{{\pm},r}\, (r\in Z\mp{\eta\over 2})$.

The equivalence of the algebras which differ by the value of the 
parameter $\eta$ can be shown by expressing the 
generators of the algebras twisted by $\eta$ 
in terms of the generators of the $\eta=0$ (Ramond) 
algebra through the relations given by Eq.(\ref{Auto}) 
with the choice of $\alpha(z)={\rm i}\eta\log z$:
\bea
L_{\rm R}(z)&=&L(z)-{\eta\over z}T^0(z)+{k\eta^2\over 4z^2}\ ,\nonumber\\ 
T^0_{\rm R}(z)&=& T^0(z)-{k\eta\over 2z}\ ,\qquad
T^{\pm}_{\rm R}(z)=z^{\mp\eta}T^{\pm}(z)\ ,\nonumber\\
G^{\mp}_{\rm R}(z)&=&z^{\pm{\eta\over 2}}G^{\mp}(z)\ ,\qquad
\bar G_{{\rm R} \pm}(z)=z^{\mp{\eta\over 2}}\bar G_{\pm}(z)\ .
\label{AutoFlowR}
\eea
or by expressing the 
generators of the algebras twisted by $1-\eta$ 
in terms of the generators of the $\eta=1$ (Neveu-Schwarz) 
algebra through the relations given by Eq.(\ref{Auto}) 
with the choice of $\alpha(z)=-{\rm i}(1-\eta)\log z$:
\bea
L_{\rm NS}(z)&=&L(z)+{(1-\eta)\over z}T^0(z)
+{k(1-\eta)^2\over 4z^2}\ ,\nonumber\\ 
T^0_{\rm NS}(z)&=& T^0(z)+{k(1-\eta)\over 2z}\,\qquad
T^{\pm}_{\rm NS}(z)=z^{\pm(1-\eta)}T^{\pm}(z)\ ,\nonumber\\
G^{\mp}_{\rm NS}(z)&=&z^{\mp{(1-\eta)\over 2}}G^{\mp}(z)\ ,\qquad
\bar G_{{\rm NS} \pm}(z)=z^{\pm{(1-\eta)\over 2}}\bar G_{\pm}(z)\ .
\label{AutoFlowNS}
\eea

Since the algebra for each choice of $\eta$ is 
equivalent to each other 
through the inner automorphism mentioned above, 
we basically consider in the following the typical choices of 
$\eta=0$ and $\eta=1$, and compare their resulting consequences.
The value $\eta=0$ in Eq.(\ref{BoundC}) corresponds 
to the Ramond (R) sector with integral $r$ and $s$, 
while that of $\eta=1$ to the Neveu-Schwarz (NS) one 
with half-integral values of them.
The spectral flow between the NS and R sectors is obtained either 
by putting $\eta=1$ in Eq.(\ref{AutoFlowR}) or 
by putting $\eta=0$ in Eq.(\ref{AutoFlowNS}).

For a general value of $\eta$ 
we consider the {\it R-type} raising operators that are given 
as follows:
\bea
L_n\quad &&(n>0)\ ,\nonumber\\
T^{ i}_{n-i\eta}\quad 
&&(n>0 \quad {\rm or}\quad  i=+\  {\rm and}\  n=0)\ ,\nonumber\\
G^{ a}_{n-{1\over 2}a\eta}\quad 
&&( n> 0 \quad{\rm or}\quad  a=+\  {\rm and}\  n=0)\ ,\nonumber\\
\bar G_{a,n-{1\over 2}a\eta}\quad 
&&( n> 0 \quad{\rm or}\quad  a=+\  {\rm and}\  n=0)\ ,
\label{RaisingOpR}
\eea
where $n\in Z$, and 
we have used the following notation: 
$i\eta=(\pm\eta,0)$ for $i=(\pm,0)$ 
and $a\eta=\pm\eta$ for $a=\pm$.
The above choice corresponds to taking the raising operators 
among the generators of the Ramond ($\eta=0)$ sector 
to be the normal ones as are usually chosen, 
which can be obtained just by putting $\eta=0$ 
in Eq.(\ref{RaisingOpR}).

But we should note that for the NS sector with $\eta=1$, 
the above choice of Eq.(\ref{RaisingOpR}) amounts to 
taking the raising operators to be the {\it tilted} ones 
that are obtained by putting $\eta=1$ in Eq.(\ref{RaisingOpR}) 
rather than the ordinary raising operators 
that are commonly used for the NS sector 
and are given 
by putting $\eta=1$ in the following {\it NS-type} conditions:
\bea
L_n\quad &&(n>0)\ ,\nonumber\\
T^i_{n+i(1-\eta)}\quad 
&&(n>0 \quad {\rm or}\quad  i=+\  {\rm and}\  n=0)\ ,\nonumber\\
G^a_{n'+{a(1-\eta)\over 2}}\quad 
&&( n'\ge {1\over 2} )\ ,\nonumber\\
\bar G_{a,n'+{a(1-\eta)\over 2}}\quad 
&&( n'\ge {1\over 2} )\ ,
\label{RaisingOpNS}
\eea
where $n\in Z,\ n'\in Z+{1\over 2}$. 
This implies that the hws's usually constructed for the NS sector 
by the condition 
of being killed by all the raising operators $X_+$ 
obtained by 
putting $\eta=1$ in Eq.(\ref{RaisingOpNS}) 
do not correspond to the hws's of the R sector, 
but rather the latter should be reconstructed properly 
from the former if one started with the hws's of 
the NS sector.  
This procedure will be exemplified 
for some relevant cases later.

We take 
the Cartan subalgebra to be $\{L_0,T^0_0,k\}$, and 
the lowering operators to be the remaining generators. 
Then 
we define a highest weight representation (hwrep) of the algebras 
Eq.(\ref{Fourier}) to be one containing 
a highest weight state (hws) vector 
$|h,\ell\big>$ such that 
\bea
L_0|h,\ell\big>&=&h|h,\ell\big>\ ,\nonumber\\
T^0_0|h,\ell\big>&=&\ell|h,\ell\big>\ ,
\label{hws}
\eea
and
\be
X_+|h,\ell\big>=0\ ,
\label{Rhws}
\ee
for all raising operators $X_+$.

Our FF representations of the 
$N=4$ SU(2)$_k$ superconformal algebra 
are constructed in terms of four bosons 
$\varphi_\alpha(z)\ (\alpha=1,2,3,4)$ and four real fermions 
forming a pair of complex fermion doublet $\gamma^a(z)$ 
and antidoublet $\bar\gamma_a(z)$\ ($a=1,2$ or $\pm$) 
under SU(2)$_k$: 
\bea
\Bigl(\gamma^a(z)\Bigr)&=&\pmatrix{
                     \gamma^1(z)   \cr
                     \gamma^2(z)   \cr
                     }
           =\pmatrix{
                     \gamma^-(z)   \cr
                     \gamma^+(z)   \cr
                     }\ ,                   \nonumber\\
\Bigl(\bar\gamma_a(z)\Bigr)&=&\Bigl(\bar\gamma_1(z),\bar\gamma_2(z) \Bigr)
           =\Bigl(\bar\gamma_+(z),\bar\gamma_-(z)  \Bigr)\ ,
\label{gammaD}
\eea
whose mode expansions are given by 
\be
\varphi_\alpha(z)=q_\alpha-{\rm i}p_\alpha\log z
+{\rm i}\sum_{n\in Z,n\not=0}{\varphi_{\alpha,n}\over n}z^{-n}
\qquad(\alpha=1,2,3,4)\ ,
\ee
and
\be
\gamma^{\eta,a}(z)=
\left\{
\begin{array}{ll}
z^{-{a(1-\eta)\over 2}}\gamma^a_{\rm NS}(z)
\equiv\sum\limits_{n'\in Z+{1\over 2}}\gamma^a_{n'}
z^{-n'-{a(1-\eta)\over 2}-{1\over 2}}
       & \mbox{NS-type}\ \ (0<\eta\le 1)   \vspace{5mm}  \\
z^{a\eta\over 2}\gamma^a_{\rm R}(z)
\equiv\sum\limits_{n\in Z}\gamma^a_{n}
z^{-n+{a\eta\over 2}-{1\over 2}}
            & \mbox{R-type}\quad (0\le\eta<1)\ ,  
\end{array}   
\right.
\label{ModeFermi}
\ee
\be
\bar\gamma^{\eta}_a(z)=
\left\{
\begin{array}{ll}
z^{-{a(1-\eta)\over 2}}\bar\gamma_{{\rm NS}a}(z)
\equiv\sum\limits_{n'\in Z+{1\over 2}}\bar\gamma_{a,n'}
z^{-n'-{a(1-\eta)\over 2}-{1\over 2}}
       & \mbox{NS-type}\ \ (0<\eta\le 1)   \vspace{5mm}  \\
z^{a\eta\over 2}\bar\gamma_{{\rm R}a}(z)
\equiv
\sum\limits_{n\in Z}\bar\gamma_{a,n}
z^{-n+{a\eta\over 2}-{1\over 2}}
            & \mbox{R-type}\quad (0\le\eta<1)\ .  
\end{array}   
\right.
\label{ModeBarFermi}  
\ee
In the following we mostly consider the R case ($\eta=0$) only 
instead of the R-type case ($0\le \eta <1$) 
for clarity and simplicity.

The commutators for the Fourier modes are 
\bea
&&[\varphi_{\alpha,m},\varphi_{\beta,n}]=m\delta_{\alpha\beta}
\delta_{m+n,0}\ ,\qquad
[q_\alpha,p_\beta]={\rm i}\delta_{\alpha\beta}\ ,\nonumber\\
&&[\varphi_{\alpha,n},q_\beta]=[\varphi_{\alpha,n},p_\beta]=0\ ,
\label{CommutatorB}
\eea
for the bosons, while for the fermions they are 
for the NS-type case ($m',n'\in Z+{1\over 2}$) 
\bea
&&\{\gamma^a_{m'},\bar\gamma_{b,n'}\}
={\delta^a}_b \delta_{m'+n',0}
=\delta_{a+b,0}\delta_{m'+n',0}\ ,\nonumber\\
&&\{\gamma^a_{m'},\gamma^b_{n'}\}
=\{\bar\gamma_{a,m'},\bar\gamma_{b,n'}\}
=0\ ,
\label{CommutatorF1}
\eea
or for the R-type case ($m,n\in Z$) 
\bea
&&\{\gamma^a_{m},\bar\gamma_{b,n}\}
={\delta^a}_b \delta_{m+n,0}
=\delta_{a+b,0}\delta_{m+n,0}\ ,\nonumber\\
&&\{\gamma^a_{m},\gamma^b_{n}\}
=\{\bar\gamma_{a,m},\bar\gamma_{b,n}\}
=0\ .
\label{CommutatorF2}
\eea

The corresponding propagators for the boson fields are given by
\be
\big<\varphi_\alpha(z)\partial\varphi_\beta(w)\big>
=\big<\partial\varphi_\beta(w)\varphi_\alpha(z)\big>
= {\delta_{\alpha \beta}\over z-w}\ ,
\label{PropagatorB}
\ee
while those for the fermions are given by
\bea
\big<\bar\gamma^\eta_a(z)\gamma^{\eta,b}(w)\big>
&=&-\big<\gamma^{\eta,b}(w)\bar\gamma^\eta_a(z)\big>   
                 \nonumber    \\
\noalign{\vskip 0.2cm}
&=&\left\{
          \begin{array}{ll} 
\displaystyle{
{{\delta_a}^b\over z-w}\Bigl({w\over z}\Bigr)^{a(1-\eta)\over 2}
}   
               & \mbox{ for NS-type$\ (0<\eta\le 1)$} \vspace{5mm}  \\ 
\displaystyle{
{{\delta_a}^b\over z-w}{{z+w}\over 2\sqrt{zw}}
}
             & \mbox{for R\qquad\quad$\ (\eta=0)$}\ ,   
         \end{array}
    \right.
\label{PropagatorF}
\eea
where we have defined the {\it NS-type} vacuum 
$|0\big>$ as
\bea
\varphi_{\alpha,n}|0\big>=p_\alpha|0\big>=0
\quad\qquad\qquad    
&&(n>0)\ ,\nonumber\\
\gamma^a_{n'}|0\big>=
\bar\gamma_{a,n'}|0\big>=0\quad  
&&(n'\ge {1\over 2})\ ,
\label{NSVacuum}
\eea
while for the zero modes of the  R sector ($\eta=0$) 
we have used the following definition of normal-ordering
~\cite{DiVecchia1,DiVecchia2,DiVecchia3,GoddardOlive}:
\be
\bar\gamma_{a,0}\gamma^b_0=
{}_\circ^\circ\bar\gamma_{a,0}\gamma^b_0{}_\circ^\circ
+\big<\bar\gamma_{a,0}\gamma^b_0\big>\ ,
\label{RZero-mode}
\ee
where
\be
{}_\circ^\circ\bar\gamma_{a,0}\gamma^b_0{}_\circ^\circ\equiv
{1\over 2}[\bar\gamma_{a,0},\gamma^b_0]\ ,\qquad
\big<\bar\gamma_{a,0}\gamma^b_0\big>\equiv
{1\over 2}\{\bar\gamma_{a,0},\gamma^b_0\}={1\over 2}{\delta_a}^b\ .
\label{RNormalOrder}
\ee
As to the non-zero modes ${}_\circ^\circ\ {}_\circ^\circ$ 
is defined to reduce 
to the usual definition $:\ :$ of normal-ordering 
where creation operators stand to the left of annihilation operators 
with appropriate sign factors.

As will be shown later, 
our formalism with the NS-type vacuum $|0\big>$ 
allows one to obtain the 
generic expression for the conformal weights 
which is valid top from the NS ($\eta=1$) sector further down to 
the R ($\eta=0$) sector, reproducing the 
right value ${1\over 16}\times 4={1\over 4}$ of the 
well-known additional constant for the Ramond conformal weight 
at the point $\eta=0$.
Note, however, that the expressions Eq(\ref{PropagatorF}) 
for the fermion propagators are not continuously connected at 
$\eta=0$. 
This is simply because the two definitions of the normal ordering 
for the NS-type sector and that for the R sector are discontinuous 
due to the presence of the zero modes in the latter.   
We believe that what we have presented is the best one can do for 
the generic treatment for the conformal weights. 

Now the FF representations are given 
with the parameters 
\be
\tau\equiv\sqrt{2\over \hat k+2}\ ,\qquad
k\equiv\hat k+1\ ,\qquad\kappa\equiv{\rm i\over 2}k\tau \ .
\label{Parameter1}
\ee
as follows:
\vspace{-3mm}
\begin{flushleft}
{\it SU(2)$_k$ Kac-Moody currents}
\end{flushleft}
\be
T^{i}(z)=\left\{
\begin{array}{ll}
\displaystyle{
J^{\eta, i}(z)+
{1\over 2}:\bar\gamma^\eta(z)\sigma^i\gamma^\eta(z):
-\delta_{i,0}{(1-\eta)\over 2z} 
}
   \vspace{2mm}  &   \\
\displaystyle{
=z^{-i(1-\eta)}J^i(z)-\delta_{i,0}{\hat k(1-\eta)\over 2z}
}     
       \vspace{2mm}   &    \\
\displaystyle{
\hskip 1cm+z^{-i(1-\eta)}
{1\over 2}:\Bigl[\bar\gamma(z)\sigma^i\gamma(z)\Bigr]_{\rm NS}:
-\delta_{i,0}{(1-\eta)\over 2z}        
}
        \vspace{2mm}    &  \\
\displaystyle{
= z^{-i(1-\eta)}T^i_{\rm NS}(z)-\delta_{i,0}{k(1-\eta)\over 2z}
}
       & \mbox{NS-type}\ \ (0<\eta\le 1)   \vspace{8mm}  \\
\displaystyle{
J^i(z)+{1\over 2}{}^\circ_\circ
\Bigl[\bar\gamma(z)\sigma^i\gamma(z)\Bigr]_{\rm R}
{}^\circ_\circ
}
   \vspace{2mm}  &   \\
\displaystyle{
=T^i_{\rm R}(z)
}
                   & \mbox{R}\quad (\eta=0)\ ,  
\end{array}   
\right. 
\label{TotalCurrents}  
\ee
where
\be
J^{\eta,\pm}(z)=z^{\mp(1-\eta)}J^{\pm}(z)\ ,\qquad
J^{\eta,0}(z)=J^0(z)-{\hat k(1-\eta)\over 2z}\ ,
\label{EtaCurrents}
\ee
with~\cite{Nemes} 
\bea
J^{\pm}(z)&=&:{\rm i\over\sqrt 2}
\left(\sqrt{\hat k+2\over 2}\partial\varphi_1(z)\pm
{\rm i}\sqrt{\hat k\over 2}\partial\varphi_2(z)\right)
e^{{\pm\rm i}\sqrt{2\over \hat k}
\left(\varphi_3(z)-{\rm i}\varphi_2(z)\right)}:\ ,   \nonumber\\
J^0(z)&=&{\rm i}\sqrt{\hat k\over 2}\partial\varphi_3(z)\ .
\label{NemeshanskyCurrents}
\eea
The boundary conditions Eq.(\ref{BoundC}) and 
the spectral flow Eq.(\ref{AutoFlowNS}) for $T^i(z)$ are 
obviously valid in Eq.(\ref{TotalCurrents}) 
when we have 
\be 
p_3-{\rm i}p_2=0\ ,
\label{momcondition}
\ee
which is valid\cite{MatIshimoto} 
for any conformal state in our FF representations, 
as will be clear from our later discussion 
on vertex operators.

\begin{flushleft}
{\it Total energy-momentum tensor}
\end{flushleft}
\be
L(z)=\left\{
\begin{array}{ll}
\displaystyle{
{1\over \hat k+2}\Biggl[
\sum\limits_{i,j=\pm,0}:J^{\eta,i}(z)\eta_{ij}J^{\eta,j}(z):\Biggr]
}
    &  \vspace{2mm}      \\
\displaystyle{
\hskip 2.5cm
    -{1\over 2}\big(\partial\varphi_4(z)\big)^2
-{\rm i}\kappa\partial^2\varphi_4(z) 
}
    &  \vspace{2mm}      \\
\displaystyle{
\hskip .5cm 
+{1\over 2}:\Bigl[\partial\bar\gamma^\eta(z)\cdot\gamma^\eta(z)
-\bar\gamma^\eta(z)\cdot\partial\gamma^\eta(z)\Bigr]:
+{(1-\eta)^2\over 4z^2}
}
                  &     \vspace{2mm}  \\
\displaystyle{
=-{1\over 2}\sum\limits_{\alpha=1}^4 
:\big(\partial\varphi_\alpha(z)\big)^2:
+{\rm i}{\tau\over 2}\partial^2\varphi_1(z)
-{\rm i}\kappa\partial^2\varphi_4(z)    
}
        &  \vspace{2mm}     \\
\displaystyle{
\hskip 2.5cm 
-{1-\eta\over z}J^0(z)+{\hat k(1-\eta)^2\over 4z^2}
}
  & \vspace{2mm} \\
\displaystyle{
\hskip .5cm 
+{1\over 2}:\Bigl[\partial\bar\gamma(z)\cdot\gamma(z)
-\bar\gamma(z)\cdot\partial\gamma(z)\Bigr]_{\rm NS}:
}
  &  \vspace{2mm}  \\
\displaystyle{
\hskip 2.5cm
-{1-\eta\over z}
{1\over 2}:\Bigl[\bar\gamma(z)\sigma^0\gamma(z)\Bigr]_{\rm NS}:   
     +{(1-\eta)^2\over 4z^2} 
}
         &    \vspace{2mm}\\
\displaystyle{
=L_{\rm NS}(z)-{1-\eta\over z}T^0_{\rm NS}(z)
+{k(1-\eta)^2\over 4z^2}
}
               & \mbox{NS-type}\ \ (0<\eta\le 1)   \vspace{7mm}  \\
\displaystyle{
{1\over \hat k+2}\Biggl[
\sum\limits_{i,j=\pm,0}:J^i(z)\eta_{ij}J^j(z):\Biggr]
}
    &  \vspace{2mm}     \\
\displaystyle{
\hskip 2.5cm
-{1\over 2}\Bigl(\partial\varphi_4(z)\Bigr)^2
-{\rm i}\kappa\partial^2\varphi_4(z) 
}
    &  \vspace{2mm}     \\
\displaystyle{
\hskip .5cm
+{1\over 2}{}^\circ_\circ\Bigl[\partial\bar\gamma(z)\cdot\gamma(z)
-\bar\gamma(z)\cdot\partial\gamma(z)\Bigr]_{\rm R}{}^\circ_\circ
+{1\over 4z^2}
}
                 &     \vspace{2mm} \\  
\displaystyle{
=
-{1\over 2}\sum\limits_{\alpha=1}^4 
:\big(\partial\varphi_\alpha(z)\big)^2:
+{\rm i}{\tau\over 2}\partial^2\varphi_1(z)
-{\rm i}\kappa\partial^2\varphi_4(z) 
}
    &     \vspace{2mm}   \\
\displaystyle{
\hskip .5cm 
+{1\over 2}{}^\circ_\circ\Bigl[\partial\bar\gamma(z)\cdot\gamma(z)
-\bar\gamma(z)\cdot\partial\gamma(z)\Bigr]_{\rm R}{}^\circ_\circ
+{1\over 4z^2}
}
    &     \vspace{2mm}   \\
\displaystyle{
=L_{\rm R}(z)
}
                 & \mbox{R}\quad (\eta=0)\ .  
\end{array}   
\right.
\label{TotalEM}  
\ee
where the spectral flow  Eq.(\ref{AutoFlowNS}) for $L(z)$
is found to hold 
by use of Eq.(\ref{TotalCurrents}).
\begin{flushleft}
{\it N=4 supercurrents}
\end{flushleft}
\be
G^{a}(z)=\left\{
\begin{array}{ll}
\displaystyle{
\gamma^{\eta,a}(z){\rm i}\partial\varphi_4(z)
-2\kappa\partial\gamma^{\eta,a}(z)
}
                 &    \vspace{2mm} \\
\displaystyle{
\hskip.5cm
-{\rm i}\tau J^{\eta,i}(z)\eta_{ij}\Bigl(\sigma^j\gamma^\eta(z)\Bigr)^a
}                     & \vspace{2mm}   \\
\displaystyle{
\hskip.5cm +\ {\rm i}\tau\Bigl[
:\Bigl(\bar\gamma^\eta(z)\cdot\gamma^\eta(z)\Bigr)\gamma^{\eta,a}(z):
-{a(1-\eta)\over 2z}\gamma^{\eta,a}(z)\Bigr]
}
    &  \vspace{2mm}  \\
\displaystyle{
=z^{-{a(1-\eta)\over 2}}
\Biggl[
\gamma^a(z){\rm i}\partial\varphi_4(z)-2\kappa\partial\gamma^a(z)
\Biggr.       
}
                  &  \\
\displaystyle{
\hskip2.5cm
-{\rm i}\tau J^i(z)\eta_{ij}\Bigl(\sigma^j\gamma(z)\Bigr)^a 
}
                                       &  \\
\displaystyle{
\hskip2.5cm +\ {\rm i}\tau\Biggl.
:\Bigl(\bar\gamma(z)\cdot\gamma(z)\Bigr)\gamma^a(z):\Biggr]_{\rm NS} 
}
                                &              \vspace{2mm}  \\
\displaystyle{
=z^{-{a(1-\eta)\over 2}}G^a_{\rm NS}(z)
}
     & \mbox{NS-type}\ \ (0<\eta\le 1)   \vspace{7mm}  \\
\displaystyle{
\Biggl[
\gamma^a(z){\rm i}\partial\varphi_4(z)-2\kappa\partial\gamma^a(z)
-{\rm i}\tau J^i(z)\eta_{ij}\Bigl(\sigma^j\gamma(z)\Bigr)^a \Biggr.
}
                                       &  \\
\displaystyle{
\hskip2cm +\ {\rm i}\tau\Biggl.
{}^\circ_\circ
\Bigl(\bar\gamma(z)\cdot\gamma(z)\Bigr)\gamma^a(z)
{}^\circ_\circ\Biggr]_{\rm R} 
}
    &     \vspace{2mm}   \\
\displaystyle{
=G^a_{\rm R}(z)
}
            & \mbox{R}\quad (\eta=0)\ ,  
\end{array}   
\right.
\label{GCurrents}  
\ee
\be
\bar G_a(z)=\left\{
\begin{array}{ll}
\displaystyle{
\bar\gamma^\eta_a(z){\rm i}\partial\varphi_4(z)
-2\kappa\partial\bar\gamma^{\eta}_a(z)
}
                 &    \vspace{2mm} \\
\displaystyle{
\hskip.5cm
+{\rm i}\tau J^{\eta,i}(z)\eta_{ij}\Bigl(\bar\gamma^\eta(z)\sigma^j\Bigr)_a
}
                                   & \vspace{2mm}   \\
\displaystyle{
\hskip.5cm -\ {\rm i}\tau\Bigl[
:\Bigl(\bar\gamma^\eta(z)\cdot\gamma^\eta(z)\Bigr)\bar\gamma^{\eta}_a(z):
+{a(1-\eta)\over 2z}\bar\gamma^{\eta}_a(z)\Bigr]
}
    &  \vspace{2mm}  \\
\displaystyle{
=z^{-{a(1-\eta)\over 2}}
\Biggl[
\bar\gamma_a(z){\rm i}\partial\varphi_4(z)-2\kappa\partial\bar\gamma_a(z)
\Biggr.       
}
                  &       \\
\displaystyle{
\hskip2.5cm
+{\rm i}\tau J^i(z)\eta_{ij}\Bigl(\bar\gamma(z)\sigma^j\Bigr)_a 
}
                                       &    \vspace{2mm}  \\
\displaystyle{
\hskip2.5cm -\ {\rm i}\tau\Biggl.
:\Bigl(\bar\gamma(z)\cdot\gamma(z)\Bigr)\bar\gamma_a(z):\Biggr]_{\rm NS} 
}
                       &   \vspace{2mm}   \\   
\displaystyle{
=z^{-{a(1-\eta)\over 2}}\bar G_{{\rm NS}a}(z)
}
    & \mbox{NS-type}\ \ (0<\eta\le 1)      \vspace{7mm}  \\
\displaystyle{
\Biggl[
\bar\gamma_a(z){\rm i}\partial\varphi_4(z)-2\kappa\partial\bar\gamma_a(z)
+{\rm i}\tau J^i(z)\eta_{ij}\Bigl(\bar\gamma(z)\sigma^j\Bigr)_a \Biggr.
}
                                       &  \\
\displaystyle{
\hskip2cm -\ {\rm i}\tau\Biggl.
{}^\circ_\circ
\Bigl(\bar\gamma(z)\cdot\gamma(z)\Bigr)\bar\gamma_a(z)
{}^\circ_\circ\Biggr]_{\rm R} 
}
    &     \vspace{2mm}   \\
\displaystyle{
=\bar G_{{\rm R}a}(z)
}
            & \mbox{R}\quad (\eta=0)\ . 
\end{array}   
\right.
\label{BarGCurrents}  
\ee
where the boundary conditions Eq.(\ref{BoundC}) and 
the spectral flow Eq.(\ref{AutoFlowNS}) for 
$G^a(z)$ and $\bar G_a(z)$ are explicitly seen to be valid.

Now we consider
the following basic vertex operators\cite{Mat1,Mat2} 
\bea
V(t,j,j_0;z)&=&:e^{{\rm i}t\varphi_4(z)}V_{j,j_0}(z):\ ,\nonumber\\
V_{j,j_0}(z)&=&:e^{-{\rm i}j\tau\varphi_1(z)}
e^{{\rm i}j_0\sqrt{2\over \hat k}
\bigl(\varphi_3(z)-{\rm i}\varphi_2(z)\bigr)}:\ .
\label{Vertex}
\eea
Note that the following OPE relations hold\cite{Mat1,Mat2}:
\bea
J^0(z)V_{j,j_0}(w)&\sim&{j_0\over z-w}V_{j,j_0}(w)\ , \nonumber\\
{\sqrt 2}J^{\pm}(z)V_{j,j_0}(w)&\sim&
{-j\pm j_0\over z-w}V_{j,j_0\pm1}(w)\ .
\label{OPEJ}
\eea

A primary state with conformal dimension $h(t,j)$ and 
SU(2) spin $(j,j_0)$ in the NS sector 
is obtained 
by operating the vertex operator on the NS-type vacuum as 
\be
|h(t,j),j_0\big>{\hskip-1mm}\big>\sim V(t,j,j_0;z=0)|0\big>\ ,
\label{Primary}
\ee
where the conformal weight $h(t,j)$ in the NS sector 
is given by 
\bea
h(t,j)&\equiv& {t^2\over 2}+\kappa t+{\tau^2\over 2}j(j+1)
                               \nonumber\\
      &=&{1\over 2}(t+\kappa)^2+{\tau^2\over 2}(j+{1\over 2})^2
      +{\hat k\over 4}         \nonumber\\
      &=&{1\over 2}(t+\kappa)^2+{\tau^2\over 2}(j-{k\over 2})^2
      +j\ .
\label{Weight}
\eea
The last identity plays a crucial role in the following discussions. 

Now the primary state Eq.(\ref{Primary}), being a hws vector in the 
NS sector when $j_0=j$, also stands for a NS-type hws vector 
$|h_\eta,\ell_\eta\big>$ 
for the generators given by 
Eqs.(\ref{TotalCurrents}),(\ref{TotalEM}),(\ref{GCurrents}) and 
(\ref{BarGCurrents}). To be more explicit, we have 
\bea
\left(L_0=L_{{\rm NS},0}-(1-\eta)T^0_{{\rm NS},0}+{k(1-\eta)^2\over 4}\right)
|h_\eta,\ell_{\eta}\big>&=&h_\eta|h_\eta,\ell_\eta\big>  \nonumber\\
\left(T^0_0=T^0_{{\rm NS},0}-{k(1-\eta)\over 2}\right)|h_\eta,\ell_\eta\big>
&=&\ell_\eta|h_\eta,\ell_\eta\big>\ ,
\label{EtaHWS}
\eea
where the state $|h_\eta,\ell_\eta\big>$ is written as
\bea 
&&\biggl| 
h_\eta\equiv  h(t,j)-(1-\eta)j+{k\over 4}(1-\eta)^2,\ 
\ell_\eta\equiv  j-{k\over 2}(1-\eta) \biggr>             \nonumber\\
&&\hspace{2cm}\equiv |h(t,j),j_0=j\big>{\hskip-1mm}\big>\ 
\sim\ V(t,j,j_0=j;z=0)|0\big>\ .
\label{hwrep}
\eea
Later on we find that $h_0=h_{\rm R},\ \ell_0=\ell_{\rm R}$\ 
and $\ h_1=h_{\rm NS},\ \ell_1=\ell_{\rm NS}$, 
and that $h_\eta$ can also be rewritten as it should be like 
\be
h_\eta=h_{\rm R}+\eta \ell_{\rm R}+\eta^2{k\over 4}\ ,
\qquad\quad \ell_\eta=\ell_{\rm R}+\eta{k\over 2}\ .
\label{hflow}
\ee

Next we consider the R sector where the ground state vacua 
$|\pm,\pm;0\big>_{\rm R}$
are 
quadruply degenerate and are defined by 
\bea
\varphi_{\alpha,n}|+,+;0\big>_{\rm R}=p_\alpha|+,+;0\big>_{\rm R}=0
\qquad
&&(n>0)\ ,\nonumber\\
\gamma^a_n|+,+;0\big>_{\rm R}=\bar\gamma_{a,n}|+,+;0\big>_{\rm R}=0
\qquad
&&(n>0 \quad {\rm or} \quad a=+\  {\rm and}\  n=0)\ ,
\label{RvacuumHW}
\eea
and
\bea
|-,+;0\big>_{\rm R}&\equiv&\gamma^-_0|+,+;0\big>_{\rm R}\ ,\quad
|+,-;0\big>_{\rm R}\ \equiv\ \bar\gamma_{-,0}|+,+;0\big>_{\rm R}\ ,   
                            \nonumber\\
|-,-;0\big>_{\rm R}&\equiv&\bar\gamma_{-,0}\gamma^-_0|+,+;0\big>_{\rm R}=
\bar\gamma_{-,0}|-,+;0\big>_{\rm R}
   =-\gamma^-_0|+,-;0\big>_{\rm R}\ .
\label{Rvacua}
\eea
Let us also note that
\be
T^0_{{\rm R},0}|\pm,\pm;0\big>_{\rm R}
=\pm{1\over 2}|\pm,\pm;0\big>_{\rm R}\ ,\quad
T^{\pm}_{{\rm R},0}|\pm,\pm;\big>_{\rm R}=0\ ,\quad
T^{\mp}_{{\rm R},0}|\pm,\pm;\big>_{\rm R}
={1\over \sqrt 2}|\mp,\mp;0\big>_{\rm R}\ ,
\label{DoubletR}
\ee
and
\be
T^0_{{\rm R},0}|\mp,\pm;0\big>_{\rm R}=0\ ,\quad
T^{\pm}_{{\rm R},0}|\mp,\pm;0\big>_{\rm R}
=T^{\mp}_{{\rm R},0}|\mp,\pm;0\big>_{\rm R}=0\ .
\label{SingletsR}
\ee
Therefore we have an iso-doublet $|\pm,\pm;0\big>_{\rm R}$ and 
two iso-singlets $|\pm,\mp;0\big>_{\rm R}$.

We find that $|+,+;0\big>_{\rm R}$ is 
the highest weight Ramond (hw R)
vacuum which satisfies the hws conditions of being annihilated by 
all the raising operators given by Eq.(\ref{RaisingOpR}) with $\eta=0$.
Also we note that
\be
L_{{\rm R},0}|+,+;0\big>_{\rm R}
={1\over 4}|+,+;0\big>_{\rm R}\ ,\qquad
T^0_{{\rm R},0}|+,+;0\big>_{\rm R}={1\over 2}|+,+;0\big>_{\rm R}\ .
\label{hwsR}
\ee

A hws vector $|h,\ell\big>_{\rm R}$ in the Ramond sector 
can be constructed as follows. 
We first apply the vertex operator on the 
hw R vacuum $|+,+;0\big>_{\rm R}$  
to obtain a primary state in the R sector as
\be
\biggl
|h(t,j),j_0;\lambda=+{1\over 2}\biggr>{\hskip-2mm}\biggr>_{\rm R}\sim
V(t,j,j_0;z=0)|+,+;0\big>_{\rm R}\ .
\label{PrimaryR}
\ee
where $\lambda=\left(\pm{1\over 2}, 0^{\pm}\right)$  generically 
denotes the  $T^0_{{\rm R},0}$ eigenvalues 
of  the ground state vacua 
$|\pm,\pm;0\big>_{\rm R}$ and $|\mp,\pm;0\big>_{\rm R}$. 
Here we define for later use the primary states 
$|h(t,j),j_0;\lambda\big>{\hskip-1mm}\big>_{\rm R}$ 
with other values of $\lambda$ as
\bea
\biggl
|h(t,j),j_0;\lambda=0^+\biggr>{\hskip-2mm}\biggr>_{\rm R}
&\equiv& G^-_{{\rm R},0}
\biggl
|h(t,j),j_0;\lambda=+{1\over 2}\biggr>{\hskip-2mm}\biggr>_{\rm R}\ ,
\vspace{2mm}
\nonumber\\
\biggl
|h(t,j),j_0;\lambda=0^-\biggr>{\hskip-2mm}\biggr>_{\rm R}
&\equiv& \bar G_{{\rm R}-,0}
\biggl
|h(t,j),j_0;\lambda=+{1\over 2}\biggr>{\hskip-2mm}\biggr>_{\rm R}\ ,
\vspace{2mm}
\nonumber\\
\biggl
|h(t,j),j_0;\lambda=-{1\over 2}\biggr>{\hskip-2mm}\biggr>_{\rm R}
&\equiv& \bar G_{{\rm R}-,0}G^-_{{\rm R},0}
\biggl
|h(t,j),j_0;\lambda=+{1\over 2}\biggr>{\hskip-2mm}\biggr>_{\rm R}\ ,
\vspace{2mm}
\nonumber\\
&\equiv&-G^-_{{\rm R},0}\bar G_{{\rm R}-,0}
\biggl
|h(t,j),j_0;\lambda=+{1\over 2}\biggr>{\hskip-2mm}\biggr>_{\rm R}\ .
\label{PrimaryRQuartet}
\eea
Then we get a hws vector $|h,\ell\big>_{\rm R}$ by putting $j_0=j$ 
in Eq.(\ref{PrimaryR}):
\be
L_{{\rm R},0}|h_{\rm R},\ell_{\rm R}\big>_{\rm R}
=h_{\rm R}|h_{\rm R},\ell_{\rm R}\big>_{\rm R}\ ,
\qquad
T^0_{{\rm R},0}|h_{\rm R},\ell_{\rm R}\big>_{\rm R}
=\ell_{\rm R}|h_{\rm R},\ell_{\rm R}\big>_{\rm R}\ ,
\label{RamondHWS}
\ee
where
\bea
\biggl|h_{\rm R}\equiv h(t,j)+{1\over 4},\ 
\ell_{\rm R}\equiv j+{1\over 2}\biggr>_{\rm R}&\equiv&
\biggl
|h(t,j),j_0=j;\lambda=+{1\over 2}\biggr>{\hskip-2mm}\biggr>_{\rm R}             \vspace{2mm}
 \nonumber\\
&\sim& V(t,j,j_0=j;z=0)|+,+;0\big>_{\rm R}\ .
\label{hwrepR}
\eea

From Eq.(\ref{Weight}) we have 
\bea
h_{\rm R}\ =\ h(t,j)+{1\over 4}
      &=&{1\over 2}(t+\kappa)^2+{\tau^2\over 2}
      \left(j+{1\over 2}\right)^2
      +{\hat k\over 4}+{1\over 4}         \nonumber\\
         &=&{1\over 2}(t+\kappa)^2+{\tau^2\over 2}\ell_{\rm R}^2
                        +{k\over 4}
                        \qquad\left(\ell_{\rm R}=j+{1\over 2}\right)\ . 
\label{WeightR}
\eea
On the other hand we get the NS conformal weight $h_{\rm NS}$ 
from  Eqs.(\ref{Weight}) and (\ref{hwrep}) 
by putting $\eta=1$ as
\bea
h_{\rm NS}\ \equiv\  h(t,j)
      &=&{1\over 2}(t+\kappa)^2+{\tau^2\over 2}
      \left(j-{k\over 2}\right)^2
      +j                   \nonumber\\
      &=&{1\over 2}(t+\kappa)^2+{\tau^2\over 2}
      \left(\ell_{\rm NS}-{k\over 2}\right)^2
      +\ell_{\rm NS}\quad\qquad(\ell_{\rm NS}=j)\ .
\label{WeightNS}
\eea
Therefore the relation
\be
h_{\rm R}=h_{\rm NS}-\ell_{\rm NS}+{k\over 4}     
\label{SpecL}
\ee
holds when the equality 
\be
\ell_{\rm R}=\pm\left( \ell_{\rm NS}-{k\over 2}\right)
\label{SpecT}
\ee
is valid.
The sign ambiguity in Eq.(\ref{SpecT}) will be   
clarified shortly, 
but here it is solved if one takes into account of 
the second relation in Eq.(\ref{hflow}).
Remarkably enough we thus have found that 
the parameter space of our FF representations has 
the spectral flow built in consistently.

Here it is appropriate to make a remark on the tilted raising 
operators in  the NS-type case. 
Suppose we start with the NS sector $(\eta=1)$ adopting the 
NS-type raising operators given by  Eq.(\ref{RaisingOpNS}). 
The SU(2)$_k$ raising operators of  $T^i_{n+i(1-\eta)}$ for example 
consist of the normal ones 
$\{T^+_0,T^{\pm}_1,T^{\pm}_2,\cdots\}_{\rm NS}$ 
at $\eta=1$, whereas 
at $\eta=0$ 
they provide the tilted ones 
$\{T^-_0,T^{\pm}_1,T^{\pm}_2,\cdots\}_{\rm R}$. 
Therefore, the NS hws vector with the highest eigenvalue of 
$\{T^0_0\}_{\rm NS}$ 
at $\eta=1$ turns into the {\it tilted} R hws vector with 
the lowest eigenvalue of  the $\eta$-twisted 
$\{T^0_0\}_{\rm R}$ at $\eta=0$.
The spectral flow top from the $\eta=1$ point further down to 
the $\eta=0$ point causes the reflection of the $T^0_0$ eigenvalues.

The reconstruction from the tilted R hws vector 
to provide the normal R hws vector is achieved 
by operating $\{T^+_{-1}\}_{\rm NS}$ 
on the normal NS hws vector $(k-2\ell_{\rm NS})$ times
as 
\bea
&&\biggl|h_{\rm R}=h_{\rm NS}-\ell_{\rm NS}+{k\over 4},\ 
\ell_{\rm R}={k\over 2}-\ell_{\rm NS}\biggr>
\nonumber\\
&&\hspace{2cm}
=\Big(T^+_{{\rm NS},-1}\Big)^{(k-2\ell_{\rm NS})}
\biggl|h_{\rm R}=h_{\rm NS}-\ell_{\rm NS}+{k\over 4},\ 
\ell_{\rm R}=\ell_{\rm NS}-{k\over 2}\biggr>\ .
\label{normalR}
\eea
where we have presumed that $(k-2\ell_{\rm NS})$ is a non-negative 
integer. 
This point will later be discussed in detail in connection with 
unitary representations.
Eq.(\ref{normalR}) also clarifies 
how the $\pm$ sign in Eq.(\ref{SpecT}) comes about 
in the spectral flow.

Next we proceed to the discussion of unitartiy conditons. 
First consider the following commutation relations 
\be
\left[T^+_{-m},T^-_m\right]=T^0_0-{k\over 2}m\ ,\qquad
m=1,2,\cdots\ .
\label{TComm}
\ee
Sandwich them between the NS hws\  
$|h_{\rm NS}=h(t,j),\ell_{\rm NS}=j\big>$ or 
the R hws\  
$|h_{\rm R}=h(t,j)+{1\over 4},\ell_{\rm R}=j+{1\over 2}\big>_{\rm R}$, 
then we obtain 
$\ell_{\rm NS}=j\le {k\over 2}m$ or 
$\ell_{\rm R}=j+{1\over 2}\le {k\over 2}m$. 
We therefore get the constraints:
\bea
{\rm NS}&:&  \ell_{\rm NS}=j\le {k\over 2}\ ,
\nonumber\\
{\rm R}&:& \ell_{\rm R}=j+{1\over 2}\le {k\over 2}\ .
\label{Maxofell}
\eea
Similarly sandwich the NS commutation relation 
$\left\{G^-_{1\over 2},\bar G_{+,-{1\over 2}}\right\}
=2L_0-2T^0_0$ or the R commutation relation 
$\left\{G^-_0,\bar G_{+,0}\right\}=2L_0-{k\over 2}$ 
between the corresponding hws's, then we get 
\bea
{\rm NS}&:& h_{\rm NS} \ge \ell_{\rm NS}\ ,  \nonumber\\
{\rm R}&:&  h_{\rm R}\ge {k\over 4}\ .
\label{Minh}
\eea

From SU(2) theory we know that 
the unitarity condition generally requires 
$\ell_{\rm NS}$ or $\ell_{\rm R}$ to be 
non-negative intergers or half-integers. 
Consequently, unitary representations also require $k$ to be 
non-negative integers:$\ k=0,1,2,\cdots$. 
Therefore we have 
from unitarity 
\bea
{\rm NS}&:&  \ell_{\rm NS}=j=0,{1\over 2},1,\cdots,{k\over 2}\ ,
\vspace{1cm}
\nonumber\\
{\rm R}&:& \ell_{\rm R}=j+{1\over 2}=0,{1\over 2},1,\cdots,{k\over 2}\ .
\label{ellvalues}
\eea

It is well known and can be seen from the Virasoro commutator with 
$m=-n=1$ in Eq.(\ref{Fourier}) 
that the conformal weight $h(t,j)$ should be real and non-negtive 
from unitarity. 
Suppose that we have a complex momentum 
\be 
t=u+{\rm i}v
\label{momt}
\ee
for the $\varphi_4$ field. 
Then the conformal weights of the hws's for the NS and R secotors are 
expressed as 
\bea
{\rm NS}:\ h_{\rm NS}&=& h(u+{\rm i}v,j)   \vspace{2mm} \nonumber\\
          &=& {u^2\over 2}-{1\over 2}\left(v+{k\over 2}\tau\right)^2
-{\rm i}u\left(v+{k\over 2}\tau\right)
          +{\tau^2\over 2}\left(j-{k\over 2}\right)^2+j\ ,
                              \vspace{3mm}   \nonumber\\
{\rm R}:\ \ h_{\rm R}&=&h(u+{\rm i}v,j)+{1\over 4}  
                                           \vspace{2mm}  \nonumber\\
          &=& {u^2\over 2}-{1\over 2}\left(v+{k\over 2}\tau\right)^2
-{\rm i}u\left(v+{k\over 2}\tau\right)
          +{\tau^2\over 2}\left(j+{1\over 2}\right)^2+{k\over 4}\ . 
\label{Complext}   
\eea
where the identity in Eq.(\ref{Weight}) has been fully used. 
The unitarity requires that 
\be 
{\rm i}u\left(v+{k\over 2}\tau\right)=0\ ,\quad
{\rm that\  is,}\quad {\rm either}\ \ u=0\ \ {\rm or}\ \ 
v+{k\over 2}\tau=0\ ,
\label{uv}
\ee
and that, when $u=0$ we have to have 
\bea
{\rm NS}&:& -\left({k\over 2}-j\right)\tau\ \le\ v+{k\over 2}\tau\ 
\le\  \left({k\over 2}-j\right)\tau\ ,
\nonumber\\
{\rm R}&:& -\left(j+{1\over 2}\right)\tau\ \le v+{k\over 2}\tau\ 
\le\ \left(j+{1\over 2}\right)\tau\ ,
\label{v}
\eea
from Eq.(\ref{Minh}), while $u$ can take any real value when 
$v+{k\over 2}\tau=0$. 

Therefore, the reality and the bounds Eq.(\ref{Minh}) of the 
conformal weights for the NS and R sectors 
already provide the strong constraints on the allowed values 
taken by the complex momentum $t$. 
Fig.1 illustrates the allowed region of the values over which 
$t=u+{\rm i}v$ sweeps over for the unitary represetations 
in each sector.

As was already discussed 
by Eguchi and Taormina\cite{EguchiT12}, and Yu\cite{Yu}, 
in contrast to the massive representations 
with 
$h_{\rm NS}>\ell_{\rm NS}$ or $h_{\rm R}>{k\over 4}$,  
the massless representations with 
the lowest conformal weights 
$h_{\rm NS}=\ell_{\rm NS}$ or $h_{\rm R}={k\over 4}$ 
are special in that 
new singular vectors occur when 
$\ell_{\rm NS}=j={k\over 2}$ or $\ell_{\rm R}=j+{1\over 2}=0$. 
At those values of $\ell$ the non-negativeness of 
the Kac-determinants requires that 
the conformal weights are restricted to the lowest values possible 
given above. 
Otherwise the Kac-determinants become negative, thus 
the unitarity is violated.
Cosequently, we have two classes of the unitary representations of 
the $N=4$ SU(2)$_k$ algebra:
\begin{flushleft}
(A) {\it Massive representations}
\end{flushleft}
\be
\left\{
\begin{array}{lll}
\displaystyle{
h_{\rm NS}>\ell_{\rm NS}\ ,  }  & 
\displaystyle{
\ell_{\rm NS}=j=0,{1\over 2},1,\cdots,{k-1\over 2}\ , }
& \mbox{NS sector}    \vspace{2mm}      \\
\displaystyle{
h_{\rm R}>{k\over 4}\ ,  }   &
\displaystyle{
\ell_{\rm R}=j+{1\over 2}={1\over 2},1,{3\over 2},\cdots,{k\over 2}\ ,  }
& \mbox{R sector}     
\end{array}   
\right.
\label{Massive}  
\ee
\vspace{1cm}
\begin{flushleft}
(B) {\it Massless representations}
\end{flushleft}
\be
\left\{
\begin{array}{lll}
\displaystyle{
h_{\rm NS}=\ell_{\rm NS}\ ,  }  & 
\displaystyle{
\ell_{\rm NS}=j=0,{1\over 2},1,\cdots,{k-1\over 2},{k\over 2}\ , }
& \mbox{NS sector}    \vspace{2mm}      \\
\displaystyle{
h_{\rm R}={k\over 4}\ ,  }   &
\displaystyle{
\ell_{\rm R}=j+{1\over 2}=0,{1\over 2},1,{3\over 2},\cdots,{k\over 2}\ ,}
& \mbox{R sector}     
\end{array}   
\right.
\label{Massless}  
\vspace{5mm}
\ee

As presented above, 
our simple arguments on the reality of the conformal weights 
expressed by the FF parameters 
have confirmed all these results, 
except the one that the conformal weights of the hws's 
in each sector 
are restricted 
to the massless values 
$h_{\rm NS}=\ell_{\rm NS}$ or $h_{\rm R}={k\over 4}$ 
when 
$\ell_{\rm NS}=j={k\over 2}$ or $\ell_{\rm R}=j+{1\over 2}=0$. 
The last result can only be drawn from the 
sub-determinant calculations. 
Fot the NS sector it was concluded by Yu\cite{Yu} 
from the non-negativeness of the sub-determinants 
that when $\ell_{\rm NS}={k\over 2}$, the region 
$h_{\rm NS}>\ell_{\rm NS}$ is forbidden, thus 
the conformal weight of the hws being restricted to the value 
$h_{\rm NS}=\ell_{\rm NS}$. 
By use of the spectral flow of Eqs.(\ref{SpecL}) and (\ref{SpecT}), 
we can conclude that for the R sector the unitarity excludes 
the region $h_{\rm R}>{k\over 4}$ 
when $\ell_{\rm R}=0$, thus restricting the conformal weight  
to the value $h_{\rm R}={k\over 4}$.

In conclusion we have incorporated the $\eta$ phase 
of the spectral flow in the Feigin-Fuks representaions of 
the $N=4$ supercoformal algebras through the twisted fermions 
and have explicitly demonstrated that 
the spectral flow is nicely operating among the generators of 
the FF representations.
The validity of the spectral flow has also been 
investigated through the 
study of the FF parameters and 
the conditions for the unitary representaions 
have been obtained 
in the FF parameter space from the reality of the conformal weights 

whose expressions are given in terms of the FF parameters.   

\vspace{.5cm}
We would like to thank Yukitaka Ishimoto and Tsuneo Uematsu 
for the help to prepare the figure.



\vspace{5cm}
\noindent
{\large Figure Caption}
\baselineskip 16pt

\vspace{1.5cm}
\noindent
Fig.1

\noindent
Complesx $t=u+{\rm i}v$ sweeps over the values on the thick lines
for the unitary representations in the NS or R sector.

\newpage
\input epsf.sty
\pagestyle{empty}
\begin{figure}
\vspace*{-4cm}
\hspace*{-3cm}
\epsfxsize=20cm
\epsfysize=28cm
\epsffile{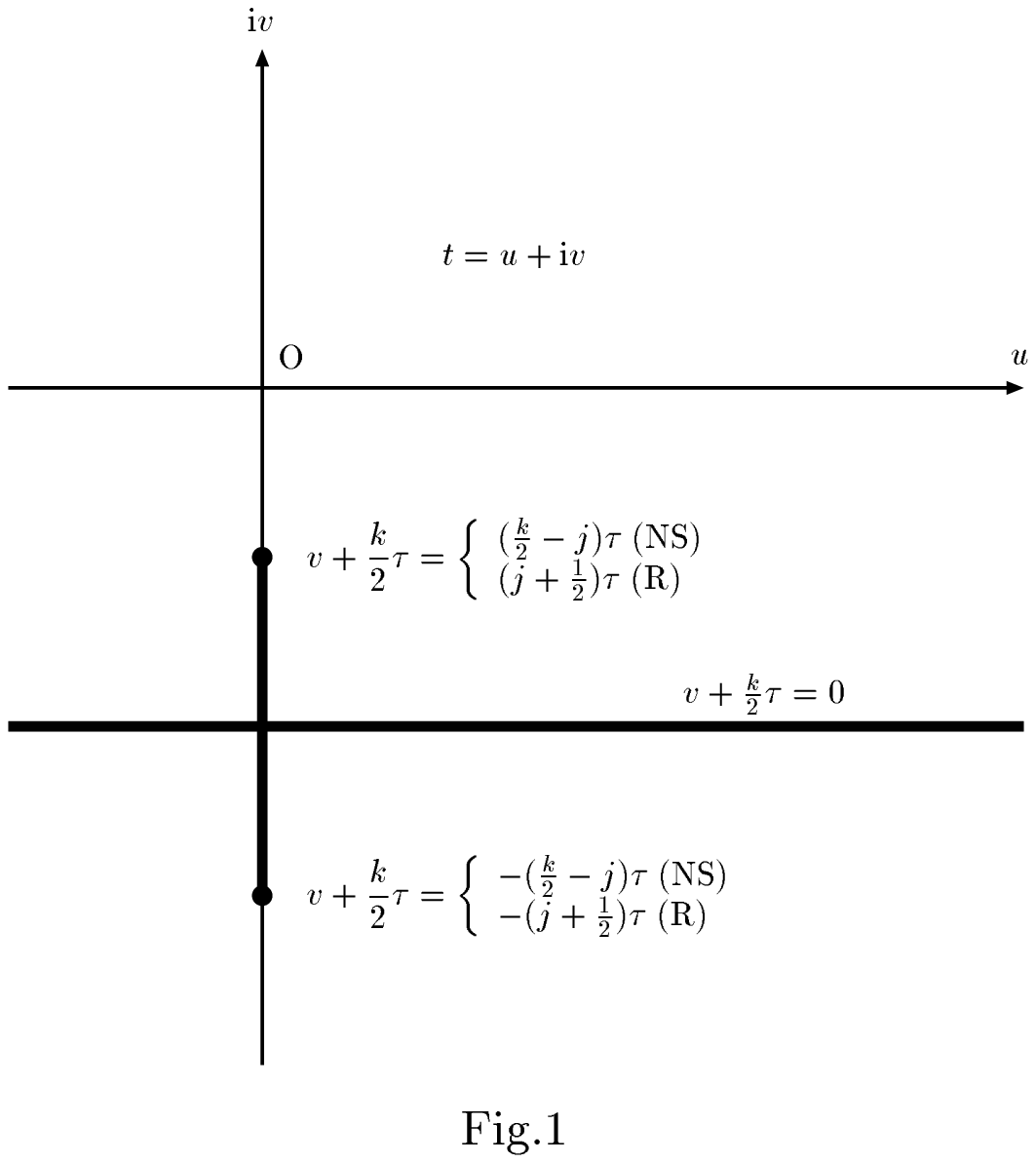}
\end{figure}

\end{document}